\documentclass[aps,PRB,reprint,superscriptaddress,preprintnumbers]{revtex4-1}
\usepackage[T1]{fontenc}
\usepackage{times}
\usepackage{graphicx,color}
\usepackage{amsfonts,amsmath,amssymb,amsbsy}
\usepackage[colorlinks=true, linkcolor=blue, citecolor=blue, urlcolor=blue]{hyperref}
\def\emph#1{\textcolor{red}{#1}}

\begin{document}

\title{Dynamics of ferromagnetic bimerons driven by spin currents and magnetic fields}

\author{Laichuan Shen}
\thanks{These authors contributed equally to this work.}
\affiliation{School of Science and Engineering, The Chinese University of Hong Kong, Shenzhen, Guangdong 518172, China}

\author{Xiaoguang Li}
\thanks{These authors contributed equally to this work.}
\affiliation{School of Science and Engineering, The Chinese University of Hong Kong, Shenzhen, Guangdong 518172, China}

\author{Jing Xia}
\affiliation{School of Science and Engineering, The Chinese University of Hong Kong, Shenzhen, Guangdong 518172, China}

\author{Lei Qiu}
\affiliation{School of Science and Engineering, The Chinese University of Hong Kong, Shenzhen, Guangdong 518172, China}

\author{\\ Xichao Zhang}
\affiliation{School of Science and Engineering, The Chinese University of Hong Kong, Shenzhen, Guangdong 518172, China}

\author{Oleg A. Tretiakov}
\affiliation{School of Physics, The University of New South Wales, Sydney 2052, Australia}

\author{Motohiko Ezawa}
\affiliation{Department of Applied Physics, The University of Tokyo, 7-3-1 Hongo, Tokyo 113-8656, Japan}

\author{Yan Zhou}
\email[Email:~]{zhouyan@cuhk.edu.cn}
\affiliation{School of Science and Engineering, The Chinese University of Hong Kong, Shenzhen, Guangdong 518172, China}

\begin{abstract}
Magnetic bimeron composed of two merons is a topological counterpart of magnetic skyrmion in in-plane magnets, which can be used as the nonvolatile information carrier in spintronic devices. Here we analytically and numerically study the dynamics of ferromagnetic bimerons driven by spin currents and magnetic fields. Numerical simulations demonstrate that two bimerons with opposite signs of topological numbers can be created simultaneously in a ferromagnetic thin film via current-induced spin torques. The current-induced spin torques can also drive the bimeron and its speed is analytically derived, which agrees with the numerical results. Since the bimerons with opposite topological numbers can coexist and have opposite drift directions, two-lane racetracks can be built in order to accurately encode the data bits. In addition, the dynamics of bimerons induced by magnetic field gradients and alternating magnetic fields are investigated. It is found that the bimeron driven by alternating magnetic fields can propagate along a certain direction. Moreover, combining a suitable magnetic field gradient, the Magnus-force-induced transverse motion can be completely suppressed, which implies that there is no skyrmion Hall effect. Our results are useful for understanding of the bimeron dynamics and may provide guidelines for building future bimeron-based spintronic devices.
\end{abstract}

\date{\today}

\keywords{bimerons, skyrmions, spintronics, micromagnetics}

\pacs{75.10.Hk, 75.70.Kw, 75.78.-n, 12.39.Dc}

\maketitle

\section{Introduction}
\label{se:Introduction}

Topologically nontrivial magnetic skyrmions have received a lot of attention, because they have small size and low depinning current, and can be used as information carriers for information storage and computing applications~\cite{Roszler_NATURE2006,Nagaosa_NNANO2013,Finocchio_JPD2016,Kang_PIEEE2016,Fert_NATREVMAT2017,ES_JAP2018,Zhou_NSR2018,Zhang_SciRep2015,Prychynenko_PRApplied2018,Nozaki_APL2019,Zhang_JPCM2020,Bogdanov_1989}.
Magnetic skyrmions have been experimentally observed in systems with bulk or interfacial Dzyaloshinskii-Moriya interaction (DMI)~\cite{Nagaosa_NNANO2013,Finocchio_JPD2016,Fert_NATREVMAT2017,ES_JAP2018}, and can be manipulated by various methods, such as electric currents~\cite{Sampaio_NatNano2013,Iwasaki_NatNano2013,Zhang_PRB2016}, spin waves~\cite{Zhang_Nanotech2015}, magnetic field gradients~\cite{Komineas_PRB2015,Wang_NJP2017,Liang_NJP2018}, magnetic anisotropy gradients~\cite{Wang_Nanoscale2018,Xia_JMMM2018,Tomasello_PRB2018,Shen_PRB2018,Ma_NanoLett2019} and temperature gradients~\cite{Kong_PRL2013,Khoshlahni_PRB2019}.
In addition, various topologically nontrivial spin textures, such as antiferromagnetic skyrmions~\cite{Zhang_SREP2016A,Barker_PRL2016,Yang_PRL2018,Shen_APL2019,Liang_PRB2019}, ferrimagnetic skyrmions~\cite{Woo_NATCOM2018}, antiskyrmions~\cite{Nayak_Nature2017}, and bimerons~\cite{Ezawa_PRB2011,Zhang_SciRep2015,Lin_PRB2015,Heo_SciRep2016,Leonov_PRB2017,Kharkov_PRL2017,Kolesnikov_SciRep2018,Chmiel_NatMater2018,Yu_Nat2018,Woo_Nat2018,Gobel_PRB2019,Fernandes_SSC2019,Murooka2019,Kim2019,Moon_arX2018,Shen_PRL2020,Lu_arXiv2020,Zarzuela_PRB2020,Zhang_Bimeron2020,Li_arXiv2020}, are also currently hot topics.

Particularly, a bimeron composed of two merons can be regarded as a counterpart of the skyrmion in in-plane magnets, which can be attained by rotating the spin texture of a skyrmion by $90^\circ$.~\cite{Gobel_PRB2019}
Therefore, the ferromagnetic (FM) bimerons share the characteristics of skyrmions, such as small size and topologically nontrivial spin structure, and they also show the transverse drift during force-driven motion, i.e., the skyrmion Hall effect~\cite{Jiang_NatPhys2017,Litzius_NatPhys2017}.
The skyrmion Hall effect may cause the skyrmion (or bimeron) to annihilate at the sample edge, which is detrimental for practical applications.
To overcome or suppress the skyrmion Hall effect, various ways have been proposed, such as adopting synthetic antiferromagnetic skyrmions~\cite{Zhang_NATCOM2016,Xia_PRApplied2019,Legrand_NatMat2019,Dohi_NATCOM2019} or applying high magnetic anisotropy in the racetrack edge~\cite{Lai_SciRep2017}. 
Magnetic bimerons, which can be found in various magnets~\cite{Kharkov_PRL2017,Zhang_Bimeron2020,Zhang_SciRep2015,Lin_PRB2015,Leonov_PRB2017,Murooka2019,Zarzuela_PRB2020,Shen_PRL2020,Li_arXiv2020,Gobel_PRB2019}, also has the potential to be used as information carriers for spintronic devices made of in-plane magnetized thin films~\cite{Zhang_SciRep2015,Gobel_PRB2019,Murooka2019}.
Recent studies on bimerons~\cite{Gobel_PRB2019,Murooka2019,Shen_PRL2020,Moon_arX2018,Zarzuela_PRB2020,Zhang_Bimeron2020,Li_arXiv2020} focus on its dynamics induced by electric currents.
However, an electric current faces the issue of Joule heating and is not applicable for insulating materials.
Therefore, it is necessary to explore alternative methods for manipulating FM bimerons effectively.
  
In this work, we report the dynamics of FM bimerons induced by spin currents and magnetic fields.
We numerically realize the simultaneous creation of two bimerons with opposite topological numbers via current-induced spin torques, and we theoretically prove that such two bimerons can coexist in a FM film with interfacial DMI.
Our results show that in addition to the spin current, a magnetic field gradient can drive a FM bimeron to motion.
Furthermore, excited by alternating magnetic field, the bimeron propagates along a certain direction, which does not show the skyrmion Hall effect when a suitable magnetic field gradient is further adopted.

\section{Model and simulation}
\label{se:Model}

\begin{figure}[t]
\centerline{\includegraphics[width=0.48\textwidth]{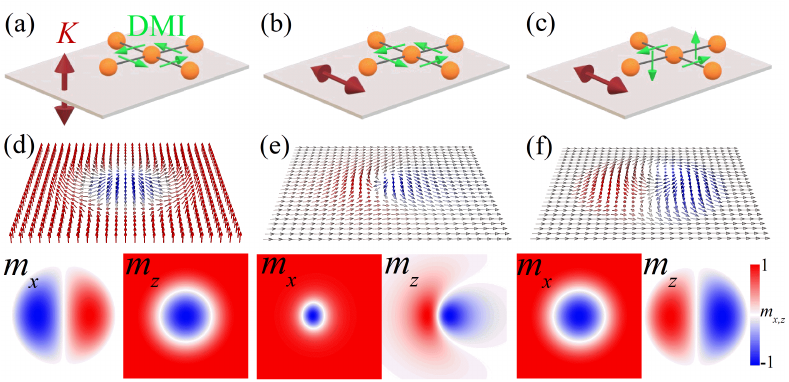}}
\caption{(a)-(c) The geometry of the magnetic atom (orange), the DMI vector (green) and the magnetic anisotropy (red). When the perpendicular magnetic anisotropy and isotropic DMI (a) are adopted, the skyrmion (d) can be a stable solution. Taking in-plane easy-axis anisotropy and isotropic DMI (b), the asymmetrical bimeron (e) is formed. If we further rotate the DMI vectors by $90^\circ$ (anisotropic DMI) (c), the bimeron has symmetrical shape (f).}
\label{FIG1}
\end{figure}

Considering a FM film with perpendicular magnetic anisotropy [Fig.~\ref{FIG1}(a)], the skyrmion [Fig.~\ref{FIG1}(d)] can be stabilized by introducing the isotropic interfacial DMI~\cite{Tomasello_SciRep2014,Rohart_PRB2013}, 
where the DMI can be induced at the ferromagnet/heavy metal (such as Ta and Pt) interface.
Here we focus on the study of a FM film with in-plane easy-axis anisotropy [Fig.~\ref{FIG1}(b)].
In such a FM system, the asymmetrical bimeron [Fig.~\ref{FIG1}(e)] is formed when the isotropic interfacial DMI is adopted.
We employ the Landau-Lifshitz-Gilbert (LLG) equation~\cite{Gilbert_IEEE2004} with the damping-like spin torque to simulate the dynamics of FM systems, which is described as
\begin{equation}
\boldsymbol{\dot{m}} = -\gamma\boldsymbol{m}\times\boldsymbol{H_{\text{eff}}} + \alpha\boldsymbol{m}\times\boldsymbol{\dot{m}} + \gamma H_{j} \boldsymbol{m}\times\boldsymbol{p}\times\boldsymbol{m}.\tag{1}
\label{eq:1}
\end{equation}
$\boldsymbol{m}$ ($= \boldsymbol{M}/M_{\text{S}}$ with saturation magnetization $M_\text{S}$) is the reduced magnetization and $\boldsymbol{\dot{m}}$ denotes the partial derivative of the magnetization with respect to time.
The damping-like spin torque $\gamma H_{j} \boldsymbol{m}\times\boldsymbol{p}\times\boldsymbol{m}$ can be produced by injecting a current into a magnetic tunnel junction or using the spin Hall effect.
$\boldsymbol{p}$ is the polarization vector and $H_{j}$ relates to the applied current density $j$, defined as $H_{j}=j\hbar\theta_{\text{SH}}/(2\mu_{0}eM_\text{S}t_{z})$ with the reduced Planck constant $\hbar$, the spin Hall angle $\theta_{\text{SH}}$, the vacuum permeability constant $\mu_{0}$, the elementary charge $e$, and the layer thickness $t_{z}$. 
$\gamma$ and $\alpha$ denote the gyromagnetic ratio and the damping constant respectively.
$\boldsymbol{H_{\text{eff}}}$ stands for the effective field obtained from the variation of the FM energy $E$,
\begin{equation}
\begin{aligned}
E=&\int{dV} \left\{ {A(\nabla\boldsymbol{m})^{\text{2}}-K(\boldsymbol{m}\cdot\boldsymbol{n})^{2}-\mu_{0} M_{\text{S}} \boldsymbol{H}\cdot\boldsymbol{m} }\right.\\
&\left.{+D[m_{z}\nabla\cdot\boldsymbol{m}-(\boldsymbol{m}\cdot\nabla)m_{z}]} \right\},
\end{aligned}
\tag{2}
\label{eq:2}
\end{equation}
where the first, second, third and fourth terms represent the exchange energy, magnetic anisotropy energy, Zeeman energy and DMI energy respectively. 
In Eq.~(\ref{eq:2}), $A$, $K$ and $D$ are the exchange constant, magnetic anisotropy constant and DMI constant respectively.
$\boldsymbol{n}=\boldsymbol{e_{x}}$ stands for the direction of the anisotropy axis, and $\boldsymbol{H}$ is the applied magnetic field. 
Note that the thermal fluctuation and dipole-dipole interaction are not taken into account.

To obtain the bimeron mentioned above, the presence of in-plane magnetic anisotropy in materials (such as CoFeB~\cite{Kipgen_JMMM2012}) is essential. On the other hand, the out-of-plane spin configurations exist in the bimeron, and they will increase the system energy if the in-plane anisotropy is considered. By introducing other energies, such as the DMI energy, which lends to spin canting, the energy increase due to the anisotropy can be compensated, so that the bimeron can be formed in a FM film with DMI and in-plane easy-axis anisotropy (the stability diagram of the bimeron is shown in Fig. S1 of Ref.~\cite{Shen_SI}). Similar to DMI, the frustrated exchange interaction can also bring the spin canting~\cite{Gobel_PRB2019}, so that the bimeron can be stabilized in frustrated FM systems~\cite{Zhang_Bimeron2020}. In addition to ferromagnets, the bimeron is a stable solution in antiferromagnets in the presence of DMI and in-plane anisotropy~\cite{Shen_PRL2020,Li_arXiv2020}. Note that the shape of bimerons depends on the DMI. Taking isotropic DMI (see Fig.~\ref{FIG1}), i.e., the DMI vectors are in-plane and DMI energy constant $D_{x} = D_{y}$, the formed bimeron has asymmetrical shape~\cite{Li_arXiv2020}. If we rotate the DMI vectors by $90^\circ$ [see Fig.~\ref{FIG1}(c)]~\cite{Gobel_PRB2019}, the bimeron shape will be symmetrical [see Fig.~\ref{FIG1}(f)]~\cite{Shen_PRL2020}. The dynamics of symmetrical and asymmetrical FM bimerons induced by alternating magnetic fields will be discussed later, while for the bimeron in antiferromagnet, it is difficult to excite its dynamics by a magnetic field.

\section{Spin current-induced creation of bimerons}
\label{se:Creation}

\begin{figure}[t]
\centerline{\includegraphics[width=0.48\textwidth]{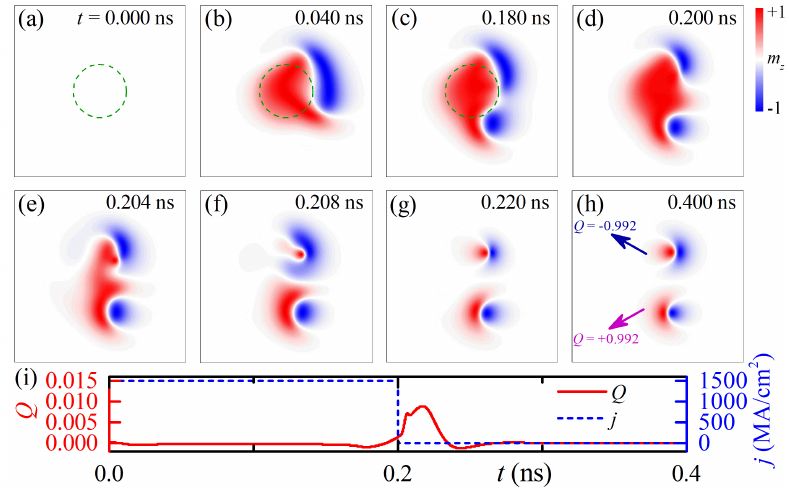}}
\caption{(a)-(h) The time evolution of the magnetization $m_{z}$ induced by a spin-polarized current with the polarization vector $\boldsymbol{p} = \boldsymbol{e_{z}}$, where the damping-like spin torque is taken into account and the color represents the value of $m_{z}$. (i) The evolution of the topological number $Q$ and the injected current density $j$. In our simulations, the current of $j=1500$ MA/cm$^{2}$ is injected in the central circular region with diameter of $30$ nm [see green lines in Figs. (a)-(c)] and we adopt the following parameters, $A=15$ pJ/m, $K=0.8$ MJ/m$^{3}$, $D=4$ mJ/m$^{2}$, $M_\text{S}=580$ kA/m, $\gamma=2.211 \times 10^{5}$ m/(A s) and $\theta_{\text{SH}}=0.2$. Here we take the damping $\alpha=0.5$, which is a realistic value in the ultra-thin FM layer grown on a heavy metal~\cite{Litzius_NatPhys2017}. The mesh size of $0.3 \times 0.3 \times 0.5$ nm$^{3}$ is used to discretize the FM film with the size $120 \times 120 \times 0.5$ nm$^{3}$. Figs. (a)-(h) only show the $m_{z}$ in the $96 \times 96$ nm$^{2}$ plane.}
\label{FIG2}
\end{figure}

Creating bimerons is the foundation for their practical applications.
Here we use a spin current to create the bimerons via damping-like spin torques.
As shown in Fig.~\ref{FIG2}(a), the initial state is the FM ground state. When the current pulse of $j = 1500$ MA/cm$^{2}$ and $\boldsymbol{p} = \boldsymbol{e_{z}}$ is injected into the central circular region with diameter of 30 nm, the magnetization in the circular region will be flipped towards the direction of the polarization vector $\boldsymbol{p}$, as shown in Fig.~\ref{FIG2}(b).   
At $t = 0.2$ ns [Fig.~\ref{FIG2}(d)], the current is switched off, and then the magnetic texture is relaxed [see Figs.~\ref{FIG2}(e)-(h)].
Figure~\ref{FIG2}(h) shows that after the relaxation, two bimerons are simultaneously generated.
In addition, we calculate the time evolution of the topological number $Q=-1/(4\pi)\int{dxdy}[\boldsymbol{m}\cdot(\partial_{x}\boldsymbol{m}\times\partial_{y}\boldsymbol{m})]$~\cite{Barker_PRL2016,Lin_PRB2015,Tretiakov_PRB2007}, showing that for the two bimerons created here, the sum of their topological numbers equals to zero [see Fig.~\ref{FIG2}(i)].
This indicates their opposite topological numbers, which is also confirmed in Fig.~\ref{FIG2}(h).
A pair of bimerons with opposite $Q$ shown in Fig.~\ref{FIG2}(h) can be separated into two independent bimerons when a suitable spin current is applied, as they have opposite drift directions (see Fig.~\ref{FIG4}).
Note that an isolated bimeron with positive $Q$ will be created when $j = +1000$ MA/cm$^{2}$, as shown in Fig. S2 of Ref.~\cite{Shen_SI}.
If the sign of current is changed, i.e., $j = -1000$ MA/cm$^{2}$, the created bimeron has negative $Q$.  

\begin{figure}[t]
\centerline{\includegraphics[width=0.48\textwidth]{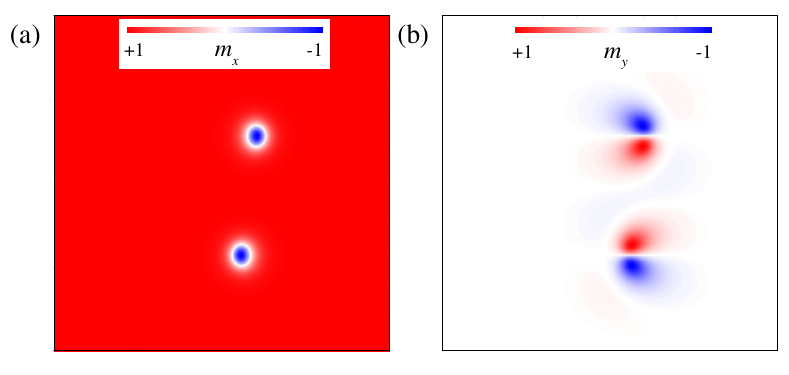}}
\caption{The components $m_{x}$ (a) and $m_{y}$ (b) of the magnetization for this case of Fig.~\ref{FIG2}(h).}
\label{FIG3}
\end{figure}

In order to analyze such a phenomenon, i.e., two bimerons with opposite signs of $Q$ are stabilized in a FM film with the same background (see Supplementary Movie 1),~\cite{Moon_arX2018,Murooka2019} we present the components $m_{x}$ and $m_{y}$ of the magnetization for the case of Fig.~\ref{FIG2}(h), as shown in Fig.~\ref{FIG3}.
Figures~\ref{FIG2}(h) and~\ref{FIG3} suggest that by using the operation $[m_{x}(x,y), m_{y}(x,y), m_{z}(x,y)]$ $\to$ $[m_{x}(-x,y), -m_{y}(-x,y), -m_{z}(-x,y)]$, two bimerons created here can convert to each other.
On the other hand, the operation mentioned above will affect the spatial derivative of the magnetization, $(\partial_{x}, \partial_{y}) \to (-\partial_{x}, \partial_{y})$.
Thus, we obtain the operation for the spatial derivative of the magnetization, $(\partial_{x}m_{x}, \partial_{x}m_{y}, \partial_{x}m_{z})$ $\to$ $(-\partial_{x}m_{x}, \partial_{x}m_{y}, \partial_{x}m_{z})$ and $(\partial_{y}m_{x}, \partial_{y}m_{y}, \partial_{y}m_{z})$ $\to$ $(\partial_{y}m_{x}, -\partial_{y}m_{y}, -\partial_{y}m_{z})$.   
Taking the above operation and combining Eq.~(\ref{eq:2}), it is found that the system energy $E$ is not changed, however giving rise to a change in the sign of the topological number $Q$.
As a result, the bimerons with opposite signs of $Q$ can coexist in the FM film with in-plane anisotropy and isotropic DMI (see Fig.~\ref{FIG1}), while in a FM system with perpendicular magnetic anisotropy and isotropic DMI, the coexistence of different skyrmions with opposite topological numbers is not allowed. We note that taking the same values of parameters, the asymmetrical bimerons have a smaller size compared to the skyrmions.

\section{spin current-driven motion of bimerons}
\label{se:Motion by spin currents}

\begin{figure}[t]
\centerline{\includegraphics[width=0.48\textwidth]{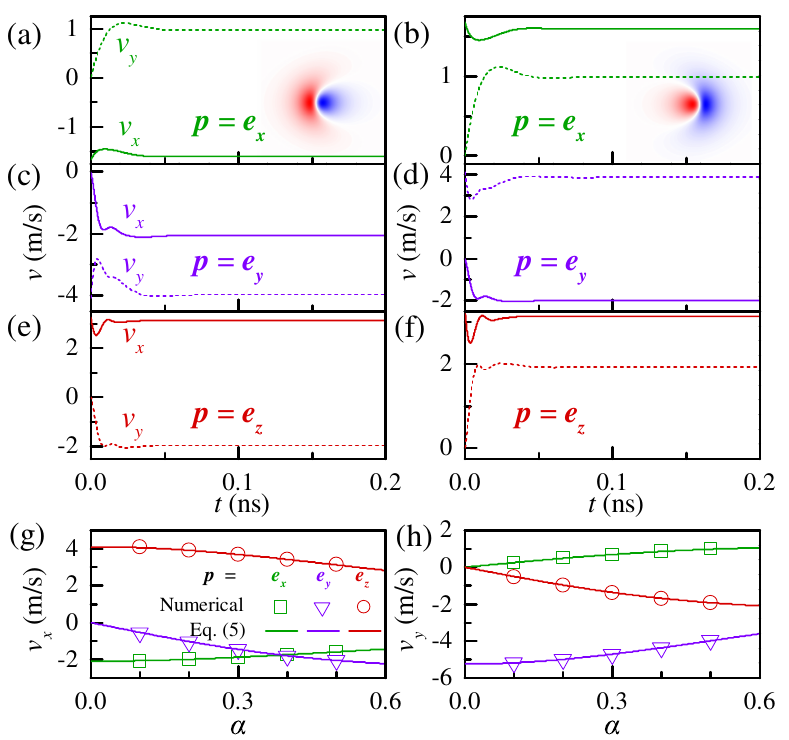}}
\caption{(a)-(f) The time evolution of the velocities ($v_{x}$, $v_{y}$) for bimerons with opposite signs of the topological number $Q$, where $j = 5$ MA/cm$^{2}$, $\theta_{\text{SH}} = 0.2$ and $\alpha = 0.5$. In addition, different polarization vectors $\boldsymbol{p} = \boldsymbol{e_{x}}, \boldsymbol{e_{y}}$ and $\boldsymbol{e_{z}}$ are adopted. The velocities $v_{x}$ (g) and $v_{y}$ (h) as functions of the damping $\alpha$ for the bimerons with positive sign of $Q$, where the numerical (symbols) and analytical (lines) results are obtained from our micromagnetic simulations and Eq.~(\ref{eq:5}) respectively. In addition to $d_{xx} \sim 15.44$ and $d_{yy} \sim 12.93$, the numerical value of ($u_{x}, u_{y}$) has been used in our calculations [Fig. S3 of Ref.~\cite{Shen_SI} shows that it is equal to (0, -13.18 nm), (33.03 nm, 0) and (0, 25.91 nm) for the cases of $\boldsymbol{p} = \boldsymbol{e_{x}}, \boldsymbol{e_{y}}$ and $\boldsymbol{e_{z}}$, respectively.].}
\label{FIG4}
\end{figure}

In addition to creating the bimerons, manipulating them is also indispensable for the application of information storage and logic devices.
We now use the spin current (instead of electric current) to manipulate the bimerons.
Taking the current density $j = 5$ MA/cm$^{2}$ and the damping $\alpha = 0.5$, the time evolution of the velocities ($v_{x}$, $v_{y}$) for bimerons with opposite signs of $Q$ is shown in Figs.~\ref{FIG4}(a)-(f), where the velocity $v_{i} = \dot{r}_{i}$ and the guiding center ($r_{x}$, $r_{y}$) of the bimeron is defined as~\cite{Komineas_PRB2015}
\begin{equation}
r_{i}=\frac {\int{dxdy}[i\boldsymbol{m}\cdot(\partial_{x}\boldsymbol{m}\times\partial_{y}\boldsymbol{m})]} {\int{dxdy}[\boldsymbol{m}\cdot(\partial_{x}\boldsymbol{m}\times\partial_{y}\boldsymbol{m})]}, i = x,y.\tag{3}
\label{eq:3}
\end{equation}

From Fig.~\ref{FIG4}, we can see that for the cases where the polarization vector $\boldsymbol{p} = \boldsymbol{e_{x}}, \boldsymbol{e_{y}}$ and $\boldsymbol{e_{z}}$, the bimerons can be driven to motion and the velocity reaches a constant value at $t = 0.2$ ns.
For the bimerons with opposite signs of $Q$, a spin current drives them to drift in the opposite directions.
Namely, their skyrmion Hall angles [$= \text{arctan}(v_{y}/v_{x})$] have opposite signs. 
The above result means that two-lane racetracks (or double-bit racetracks) can be built in order to accurately encode the data bits, where the presence of a bimeron in the top and bottom lanes is used to encode the data bits ``1'' and ``0'' respectively.~\cite{Moon_arX2018,Muller_NJP2017,Kang_IEDL2016,Lai_Spin2017}
Compared to the single-lane racetrack based on skyrmions, such a two-lane racetrack based on bimerons is robust for the data representation, as we always detect a bimeron for the data bits ``1'' and ``0''.
In the single-lane racetrack, the data bits ``1'' and ``0'' are encoded by the presence and absence of a skyrmion respectively.
However, the distance between two skyrmions may be affected by many factors, so that the number of the data bit ``0'' cannot be accurately encoded by the absence of a skyrmion.

By changing the damping constant $\alpha$, the different values of the velocities ($v_{x}$, $v_{y}$) are obtained from the numerical simulations, as shown in Figs.~\ref{FIG4}(g) and (h).
In order to test the reliability of the simulated speeds, from Eq.~(\ref{eq:1}), we derive the steady motion equation for the FM bimeron using Thiele's (or collective coordinate) approach~\cite{Thiele_PRL1973,Tveten_PRL2013,Tretiakov_PRL2008,Clarke_PRB2008}, written as
\begin{equation}
\boldsymbol{G}\times\boldsymbol{v}+\boldsymbol{F}_{\alpha}+\boldsymbol{F}_{\text{driv}} = \boldsymbol{0},\tag{4}
\label{eq:4}
\end{equation}
where $\boldsymbol{G} = (4\pi Q \mu_{0} M_{\text{S}}t_{z}/\gamma)\boldsymbol{e_{z}}$ is the gyrovector.
$\boldsymbol{F}_{\alpha}=-\alpha\mu_{\text{0}}M_\text{S}t_{z}\boldsymbol{d}\cdot\boldsymbol{v}/\gamma$ represents the dissipative force with the dissipative tensor $\boldsymbol{d}$, where the component $d_{ij}$ of the dissipative tensor is $d_{ij}=\int{dxdy}(\partial_{i}\boldsymbol{m}\cdot\partial_{j}\boldsymbol{m})$.
$\boldsymbol{F}_{\text{driv}}$ denotes the driving force, which is described as $\boldsymbol{F}_{\text{driv}}=-\mu_{0}H_{j}M_\text{S}t_{z}\boldsymbol{u}$ with $u_{i}=\int{dxdy} [(\boldsymbol{m}\times\boldsymbol{p})\cdot\partial_{i}\boldsymbol{m}]$ when the damping-like spin torque is considered.
From Eq.~(\ref{eq:4}), we obtain the steady motion speed,
\begin{equation}
\begin{pmatrix} v_{x}\\v_{y} \end{pmatrix}=\frac{\gamma \boldsymbol{L}\cdot\boldsymbol{F}_{\text{driv}}} {\mu_{0} M_{\text{S}}t_{z}[(4\pi Q)^{2}+\alpha^{2}d_{xx}d_{yy}]},\tag{5}
\label{eq:5}
\end{equation}
where $\boldsymbol{L} = \begin{pmatrix} \alpha d_{yy} & -4\pi Q\\ 4\pi Q & \alpha d_{xx} \end{pmatrix}$, and $d_{xy} \sim 0$ has been used~\cite{Shen_SI}.
Figures~\ref{FIG4}(g) and (h) show that the analytical results given by Eq.~(\ref{eq:5}) are in good agreement with the numerical simulations.
Note that for the case of $\boldsymbol{p} = \boldsymbol{e_{x}}$ (along the anisotropy axis), asymmetrical bimerons can be driven, while for a symmetrical bimeron, $\boldsymbol{F}_{\text{driv}}=\boldsymbol{0}$ is derived so that its motion speed is equal to zero.

\section{Motion of bimerons driven by magnetic field gradients}
\label{se:Motion by magnetic field gradients}

\begin{figure}[t]
\centerline{\includegraphics[width=0.48\textwidth]{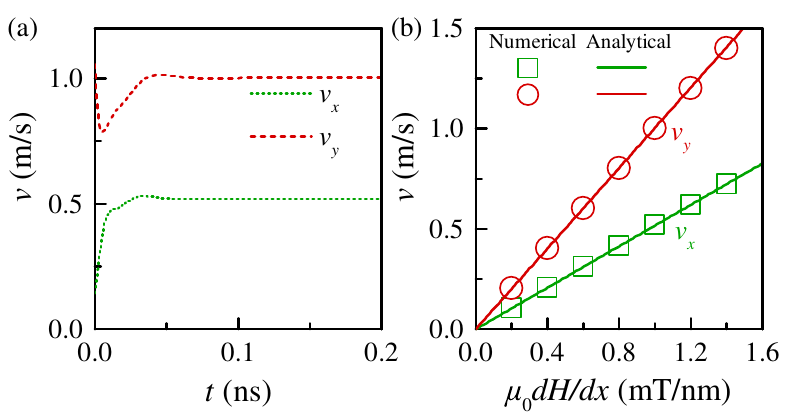}}
\caption{(a) The time evolution of the velocities ($v_{x}$, $v_{y}$) for the bimeron with positive $Q$, where the magnetic field gradient ($\boldsymbol{H} = -dH/dx \cdot x \boldsymbol{e_{x}}$ with $\mu_{0} dH/dx = 1$ mT/nm) is adopted as the driving source, and the damping $\alpha = 0.5$. (b) The velocities ($v_{x}$, $v_{y}$) at $t = 0.2$ ns as functions of the magnetic field gradient $\mu_{0} dH/dx$ for $\alpha = 0.5$, where the symbols are the results of numerical simulations and the lines are given by Eqs.~(\ref{eq:5}) and~(\ref{eq:6}).}
\label{FIG5}
\end{figure}

Using the magnetic field as a driving source is applicable in both metals and insulating materials, so it is an important manipulation method, and it is necessary to discuss the dynamics of bimeron induced by magnetic fields.
In the next sections, we focus on the study of the bimeron dynamics induced by magnetic field gradients and alternating magnetic fields.
Figure~\ref{FIG5}(a) shows the time evolution of the velocities ($v_{x}$, $v_{y}$) for the bimeron with positive $Q$, where a magnetic field gradient ($\boldsymbol{H} = -dH/dx \cdot x \boldsymbol{e_{x}}$ with $\mu_{0} dH/dx = 1$ mT/nm, where the gradient value could be feasible in experiments~\cite{Jakobi_NatNano2017}) is applied and the damping $\alpha = 0.5$.
For the case shown in Fig.~\ref{FIG5}(a), the change of bimeron size induced by the magnetic field can be ignored, and the velocities ($v_{x}$, $v_{y}$) at $t = 0.2$ ns almost reach a constant value of (0.518 m/s, 1.002 m/s).
Figure~\ref{FIG5}(b) shows that the velocities of the bimeron are proportional to the magnetic field gradient.
Due to the presence of the magnetic field gradient, the potential energy of the system changes spatially, so that a nonzero driving force will act on the FM bimeron.
We now derive the formula of the driving force induced by the magnetic field with a constant gradient. 
Applying a partial integration,~\cite{Komineas_PRB2015} the induced force is derived from $-\mu_{0}M_\text{S}t_{z}\int{\boldsymbol{H}\cdot\partial_{x}\boldsymbol{m}dxdy}$, which is described as
\begin{equation}
\boldsymbol{F}_{\text{grad}} = \mu_{0}M_\text{S}t_{z} u_{\text{H}} \frac{dH} {dx} \boldsymbol{e_{x}},\tag{6}
\label{eq:6}
\end{equation}
where $u_{\text{H}}=\int{(1-m_{x})dxdy}$ is 94.4 nm$^2$ for the parameters used here~\cite{Shen_SI}.
Substituting the above Eq.~(\ref{eq:6}) into Eq.~(\ref{eq:5}), we obtain the steady motion speed, and find that the bimeron moves towards the area of lower magnetic field, similar to the case of the skyrmion~\cite{Wang_NJP2017}.
As shown in Fig.~\ref{FIG5}(b), the results given by Eqs.~(\ref{eq:5}) and~(\ref{eq:6}) are consistent with the numerical simulations.

\section{Motion of bimerons driven by alternating magnetic fields}
\label{se:Motion by alternating magnetic fields}

\begin{figure}[t]
\centerline{\includegraphics[width=0.48\textwidth]{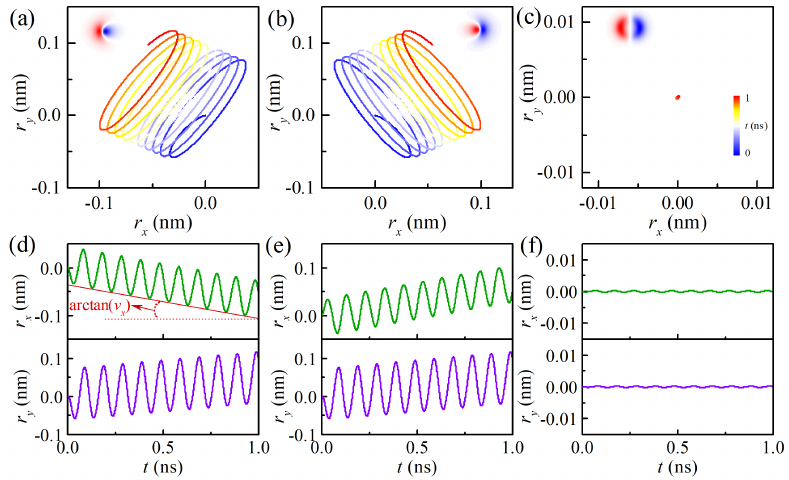}}
\caption{(a)-(c) The trajectories of bimerons induced by an alternating magnetic field $\boldsymbol{H} = H_{0}$sin($2\pi ft$)$\boldsymbol{e_{x}}$ [(a) the asymmetrical bimeron with positive $Q$; (b) the asymmetrical bimeron with negative $Q$; (c) the symmetrical bimeron with positive $Q$.], where $\mu_{0} H_{0} = 50$ mT, $f = 10$ GHz and $\alpha = 0.5$. In order to form the symmetrical bimeron, the anisotropic DMI is used in our simulation. (d)-(f) show the time evolution of the guiding center ($r_{x}$, $r_{y}$) for the cases of Figs. (a)-(c) respectively.} 
\label{FIG6}
\end{figure}

Recently, an interesting method for manipulating magnetic skyrmions, was proposed, i.e., by using an oscillating electric field and combining a static magnetic field, where the electric field can modify the magnetic anisotropy.~\cite{Yuan_PRB2019,Song_JPD2019}
Such a method is applicable in both metals and insulators. 
With the oscillating electric field alone, the skyrmion will exhibit the breathing motion (rather than propagation along a certain direction) and the spin wave excitation is symmetric.
When an in-plane static magnetic field is applied, the symmetry of the skyrmion is broken and then the spin wave excitation becomes asymmetric, resulting in nonzero net driving force.~\cite{Yuan_PRB2019,Song_JPD2019}
Thus, an oscillating electric field can drive the asymmetrical skyrmion to move along a certain direction.
In addition, the skyrmion speed reaches its maximum value when the frequency of the oscillating electric field matches the eigenfrequency of the system.~\cite{Yuan_PRB2019}

The bimerons studied here have intrinsic asymmetrical structure, so that they can be driven by the alternating field.
Figures~\ref{FIG6}(a) and (b) show the trajectories of asymmetrical bimerons driven by an alternating magnetic field $\boldsymbol{H} = H_{0}$sin($2\pi ft$)$\boldsymbol{e_{x}}$, where the applied magnetic field is uniform in space.
Indeed, the propagation of asymmetrical bimerons is induced, and for the bimerons with opposite signs of $Q$, their trajectories are essentially the same, except for their motion directions.   
For the purpose of comparison, we also calculate the motion of the symmetrical bimeron under the same magnetic field, as shown in Fig.~\ref{FIG6}(c), from which we can see that an alternating magnetic field cannot induce the symmetrical bimeron to propagation.
Note that in addition to the alternating magnetic field, the alternating magnetic anisotropy can also excite the asymmetrical bimeron to move along a certain direction, as shown in Fig. S5 of Ref.~\cite{Shen_SI}.

\begin{figure}[t]
\centerline{\includegraphics[width=0.48\textwidth]{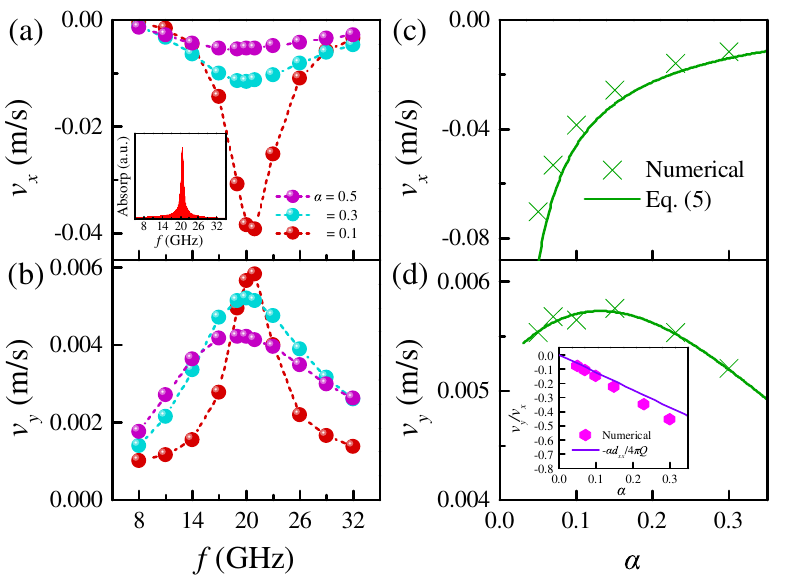}}
\caption{(a) and (b) The bimeron velocities ($v_{x}$, $v_{y}$) as functions of the frequency $f$ of the alternating magnetic field $\boldsymbol{H} = H_{0}$sin($2\pi ft$)$\boldsymbol{e_{x}}$, where $\mu_{0} H_{0} = 10$ mT, and $\alpha =$ 0.1, 0.3 and 0.5. The inset in Fig.~\ref{FIG7}(a) is the absorption spectrum. Applying a magnetic field $\boldsymbol{B_{s}} = 0.5$ mT$\cdot$sin($2\pi ft$)/($2\pi ft$)$\boldsymbol{e_{x}}$ with $f = 100$ GHz yields the average magnetization $\left \langle m_{x}(t) \right \rangle$ of the system, and then the Fourier transform of $\left \langle m_{x}(t) \right \rangle$ gives the absorption spectrum, where the damping $\alpha =$ 0.008 is used in our calculation. (c) and (d) The bimeron velocities ($v_{x}$, $v_{y}$) as functions of the damping $\alpha$, where the alternating magnetic field has the amplitude of 10 mT and frequency of 20 GHz. The inset in Fig.~\ref{FIG7}(d) shows the ratio of $v_{y}/v_{x}$ with different damping constants.} 
\label{FIG7}
\end{figure}

On the other hand, based on the time evolution of the guiding center ($r_{x}$, $r_{y}$), the propagation velocities ($v_{x}$, $v_{y}$) of the bimeron can be obtained [see Fig.~\ref{FIG6}(d)]. 
Figures~\ref{FIG7}(a) and (b) show the velocities as functions of the frequency $f$, where we take the alternating magnetic field $\boldsymbol{H} = H_{0}$sin($2\pi ft$)$\boldsymbol{e_{x}}$ with amplitude $\mu_{0} H_{0}$ of 10 mT and frequency $f$ of $8 \sim 32$ GHz.
Similar to the case of the skyrmion,~\cite{Yuan_PRB2019} the bimeron reaches its maximum speed when the frequency of alternating magnetic fields coincides with the system eigenfrequency of $\sim 20.6$ GHz [see the inset in Fig.~\ref{FIG7}(a)].
In addition, the velocities ($v_{x}$, $v_{y}$) are calculated as functions of the damping $\alpha$, as shown in Figs.~\ref{FIG7}(c) and (d), where the alternating magnetic field has the amplitude of 10 mT and frequency of 20 GHz.
To understand the results of the numerical simulations, we try to find the net driving force induced by the alternating magnetic field.
As shown in the inset of Fig.~\ref{FIG7}(d), the ratio $v_{y}/v_{x}$ obtained from numerical simulations can be described by $\alpha d_{xx}/(-4\pi Q)$.
Thus, Eq.~(\ref{eq:5}) suggests that the net driving force $\boldsymbol{F}_{\text{net}}$ is almost along the $y$ direction.
Assuming that $F_{\text{net}} = c_{1}/\alpha + c_{2} + c_{3}\alpha + c_{4}\alpha^{2}$ with $c_{1} = 0.865 \times 10^{-16}$ N, $c_{2} = 1.398 \times 10^{-16}$ N, $c_{3} = -5.081 \times 10^{-16}$ N and $c_{4} = 4.975 \times 10^{-16}$ N, the results given by Eq.~(\ref{eq:5}) agree with the numerical simulations, as shown in Figs.~\ref{FIG7}(c) and (d).
It is worth mentioning that the bimeron will annihilate, when a strong alternating magnetic field (its amplitude and frequency are 100 mT and 20 GHz respectively) is applied (see Supplementary Movie 2).
   
\section{Magnetic bimerons showing no skyrmion Hall effect}
\label{se:Magnetic bimerons}

\begin{figure}[t]
\centerline{\includegraphics[width=0.48\textwidth]{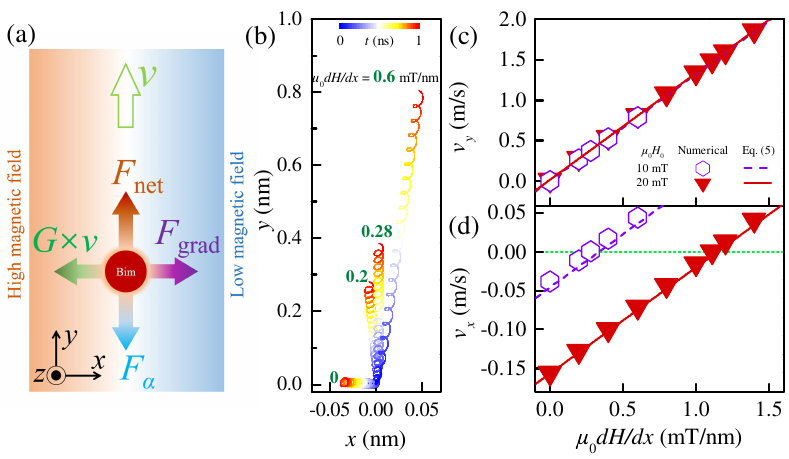}}
\caption{(a) The schematic diagram of the forces acting on the bimerons, where $\boldsymbol{F}_{\text{net}}$ and $\boldsymbol{F}_{\text{grad}}$ are the forces induced by the alternating magnetic field and magnetic field gradient respectively, and $\boldsymbol{F}_{\alpha}$ and $\boldsymbol{G} \times \boldsymbol{v}$ denote the dissipative force and Magnus force. The color of the background represents the value of the space-dependent magnetic field.
(b) The trajectories of bimerons induced by an alternating magnetic field $\boldsymbol{H} = H_{0}$sin($2\pi ft$)$\boldsymbol{e_{x}}$ with $\mu_{0} H_{0} = 10$ mT and $f = 20$ GHz, where we set the damping $\alpha =$ 0.1 and the total time $t =$ 1 ns in our simulations. In addition, the magnetic field gradient ($\boldsymbol{H} = -dH/dx \cdot x \boldsymbol{e_{x}}$ with $\mu_{0}dH/dx =$ 0, 0.2, 0.28 and 0.6 mT/nm) is applied.
(c) and (d) The bimeron velocities ($v_{x}$, $v_{y}$) as functions of the magnetic field gradient $\mu_{0}dH/dx$, where $\alpha = 0.1$, the frequency $f$ of alternating magnetic fields is 20 GHz, and the amplitudes $\mu_{0} H_{0}$ of 10 and 20 mT are adopted. The symbols and lines denote the results of numerical simulations and Eq.~(\ref{eq:5}) respectively.} 
\label{FIG8}
\end{figure}

As shown in Figs.~\ref{FIG6} and \ref{FIG7}, under the action of the alternating magnetic field, the FM bimerons show the skyrmion Hall effect due to the presence of the Magnus force ($\boldsymbol{G} \times \boldsymbol{v}$), and the motion speed is small ($< 0.1$ m/s).
Here we introduce a force induced by magnetic field gradients to compensate the Magnus force, and then achieve this purpose of overcoming or suppressing the skyrmion Hall effect, as shown in Fig.~\ref{FIG8}(a), where the direction of the net driving force $\boldsymbol{F}_{\text{net}}$ (the $y$ direction) is parallel to the racetrack.
Figure~\ref{FIG8}(b) shows that as the magnetic field gradient $\mu_{0}dH/dx$ increases, the bimeron moves faster and the skyrmion Hall effect is effectively suppressed. When $\mu_{0}dH/dx$ increases to 0.28 mT/nm, the bimeron propagates parallel to the racetrack with the speed of $\sim 0.374$ m/s, so that the bimeron will not be destroyed by touching the racetrack edge.
On the other hand, for the case of $\mu_{0}dH/dx =$ 0.28 mT/nm, the propagation of the bimeron is perpendicular to the gradient direction, so that the bimeron size is not affected by the space-dependent magnetic field.
If $\mu_{0}dH/dx >$ 0.28 mT/nm, the force $\boldsymbol{F}_{\text{grad}}$ induced by the magnetic field gradient is larger than the Magnus force ($\boldsymbol{G} \times \boldsymbol{v}$), causing that the bimeron moves towards the area of lower magnetic field [see Fig.~\ref{FIG8}(b)].

By changing the magnetic field gradient $\mu_{0}dH/dx$, the different velocities are obtained by numerical simulations, as shown in Figs.~\ref{FIG8}(c) and (d).
In our simulations, the damping $\alpha =$ 0.1, the frequency $f$ of alternating magnetic fields is 20 GHz, and the amplitudes $\mu_{0} H_{0}$ of 10 and 20 mT are adopted.
Combining the Eq.~(\ref{eq:6}) and substituting $\boldsymbol{F}_{\text{driv}} = \begin{pmatrix} F_{\text{grad}} \\ F_{\text{net}} \end{pmatrix}$ into Eq.~(\ref{eq:5}), the analytical velocities are given.
As shown in Figs.~\ref{FIG8}(c) and (d), the results given by Eq.~(\ref{eq:5}) are consistent with numerical simulations, where the values of $F_{\text{net}} \sim 9.58 \times 10^{-16}$ and $32.59 \times 10^{-16}$ N have been used for the cases of $\mu_{0} H_{0} =$ 10 and 20 mT respectively.
We now derive the critical magnetic field gradient, at which the bimeron moves without showing the skyrmion Hall effect, i.e., $v_{x} = 0$.
Based on $\boldsymbol{G} \times v_{y}\boldsymbol{e_{y}} + \boldsymbol{F}_{\text{grad}} = \boldsymbol{0}$, the critical magnetic field gradient is derived, which satisfies the following formula,
\begin{equation}
\frac{dH} {dx} = \frac{4\pi Q} {\gamma u_{\text{H}}} v_{y},\tag{7}
\label{eq:7}
\end{equation}
where $v_{y} = \gamma F_{\text{net}}/(\alpha d_{yy} \mu_{0} M_{\text{S}}t_{z})$.
As mentioned earlier, for the case of $\mu_{0} H_{0} =$ 10 mT, our numerical simulation shows that the critical magnetic field gradient is $\sim$ 0.28 mT/nm and the corresponding speed $v_{y}$ is $\sim$ 0.374 m/s.
If the amplitude $\mu_{0} H_{0}$ of the alternating magnetic field is increased to 20 mT, the critical magnetic field gradient and corresponding speed $v_{y}$ are $\sim$ 1.11 mT/nm and 1.486 m/s, respectively, as shown in Figs.~\ref{FIG8}(c) and (d).  
The above values obtained from numerical simulations obey Eq.~(\ref{eq:7}).

\section{Conclusions}
\label{se:Conclusions}

In conclusion, we analytically and numerically study the dynamics of FM bimerons induced by spin currents and magnetic fields.
Numerical simulations show that two bimerons with opposite signs of the topological numbers can be simultaneously created in a FM film via current-induced spin torques, and we prove their energy equivalence.
However, the coexistence of two skyrmions with opposite topological numbers is not allowed in a FM film with the same background. Compared to the skyrmions, the bimerons studied here have a smaller size for the same values of parameters.
The motion of bimerons induced by spin currents is also discussed, and the bimeron speed is analytically derived, which agrees well with the numerical simulations.
We point out that two-lane racetracks based on bimerons can be built in order to accurately encode the data bits, as the bimerons with opposite topological numbers can coexist and have opposite drift directions.
In addition, a magnetic field gradient can drive a bimeron towards the area of lower magnetic field.
More importantly, when only an alternating magnetic field is applied to the entire film system, the asymmetrical bimeron propagates along a certain direction.
Besides, if a suitable magnetic field gradient is further introduced, the alternating magnetic field can drive the bimeron to move at a speed of $\sim$ 1.5 m/s and the bimeron does not show the skyrmion Hall effect.
Our results are useful for understanding of the bimeron dynamics and may provide effective ways for building bimeron-based spintronic devices.

\begin{acknowledgments}
X.L. acknowledges the support by the Guangdong Basic and Applied Basic Research Foundation (Grant No. 2019A1515111110).
X.Z. acknowledges the support by the Guangdong Basic and Applied Basic Research Foundation (Grant No. 2019A1515110713), and the Presidential Postdoctoral Fellowship of The Chinese University of Hong Kong, Shenzhen (CUHKSZ).
O.A.T. acknowledges the support by the Australian Research Council (Grant No. DP200101027) and the Cooperative Research Project Program at the Research Institute of Electrical Communication, Tohoku University.
M.E. acknowledges the support by the Grants-in-Aid for Scientific Research from JSPS KAKENHI (Grant Nos. JP18H03676 and JP17K05490) and the support by CREST, JST (Grant Nos. JPMJCR16F1 and JPMJCR1874).
Y.Z. acknowledges the support by the President's Fund of CUHKSZ, Longgang Key Laboratory of Applied Spintronics, National Natural Science Foundation of China (Grant Nos. 11974298 and 61961136006), Shenzhen Fundamental Research Fund (Grant No. JCYJ20170410171958839), Shenzhen Key Laboratory Project (Grant No. ZDSYS201603311644527), and Shenzhen Peacock Group Plan (Grant No. KQTD20180413181702403).
\end{acknowledgments}



\end{document}